\documentclass[10pt,twocolumn,letterpaper]{article}

\usepackage[pagenumbers]{cvpr} 

\usepackage[dvipsnames]{xcolor}

\definecolor{cvprblue}{rgb}{0.21,0.49,0.74}
\usepackage{graphicx}
\usepackage{nicematrix,enumitem}
\usepackage{booktabs}
\usepackage{multirow}
\usepackage[tableposition=top]{caption}
\usepackage{subcaption}
\usepackage[pagebackref,breaklinks,colorlinks,citecolor=cvprblue]{hyperref}

\def\paperID{37} 
\def\confName{CVPR}
\def\confYear{2024}

\title{Enhancing Clinically Significant Prostate Cancer Prediction in T2-weighted Images through Transfer Learning from Breast Cancer}

\author{Chi-en Amy Tai \\
University of Waterloo\\
Waterloo, ON\\
{\tt\small amy.tai@uwaterloo.ca}
\and Alexander Wong \\
University of Waterloo\\
Waterloo, ON\\
{\tt\small alexander.wong@uwaterloo.ca}
}

\begin{document}
\maketitle
\begin{abstract}
In 2020, prostate cancer saw a staggering 1.4 million new cases, resulting in over 375,000 deaths. The accurate identification of clinically significant prostate cancer is crucial for delivering effective treatment to patients. Consequently, there has been a surge in research exploring the application of deep neural networks to predict clinical significance based on magnetic resonance images. However, these networks demand extensive datasets to attain optimal performance. Recently, transfer learning emerged as a technique that leverages acquired features from a domain with richer data to enhance the performance of a domain with limited data. In this paper, we investigate the improvement of clinically significant prostate cancer prediction in T2-weighted images through transfer learning from breast cancer. The results demonstrate a remarkable improvement of over 30\% in leave-one-out cross-validation accuracy.
\end{abstract}    
\section{Introduction}
\label{sec:intro}

In 2020, there were over 1.4 million new cases of prostate cancer with over 375,000 new deaths~\cite{sung2021global}. Identifying clinically significant prostate cancer is critical as it has the potential to grow and spread, thereby posing a significant risk to patient health. Overdiagnosis and overtreatment of clinically insignificant prostate cancer can lead to avoidable side effects and psychological distress for patients. Conversely, underdiagnosis of clinically significant prostate cancer delays needed treatment and compromises patient survival~\cite{futterer2015can}. 

T2-weighted (T2w) magnetic resonance imaging (MRI) images have been routinely ingrained in the diagnosis of prostate cancer as they provide detailed anatomical information that enhances the detection and localization of clinically significant prostate cancer~\cite{weinreb2016pi}. However, the interpretation of T2w images is complex and highly variable in the extreme parts of the gland~\cite{montagne2021challenge}. In response, research has been conducted on detecting clinically significant prostate cancer from MRI images using deep learning convolutional neural networks~\cite{abbasi2020detecting,yoo2019prostate}. 

Unfortunately, the training of these deep models requires extensive annotated datasets to achieve high performance, which are time-consuming and expensive to obtain, especially for medical imaging~\cite{varoquaux2022machine}. Transfer learning is a concept that has been recently introduced to address the problems of smaller datasets and leverages a model pretrained on a larger dataset from one domain (or cancer type). This approach leverages learned features from other types of cancer to improve the performance of the current use case~\cite{torrey2010transfer}. 

In this paper, we examine the efficacy of leveraging pretraining from another type of cancer, notably breast cancer, to improve the prediction of clinically significant prostate cancer for T2w images. 

\section{Methodology}
\label{sec:methodology}
\begin{figure*}
  \centering
    \includegraphics[width=\textwidth]{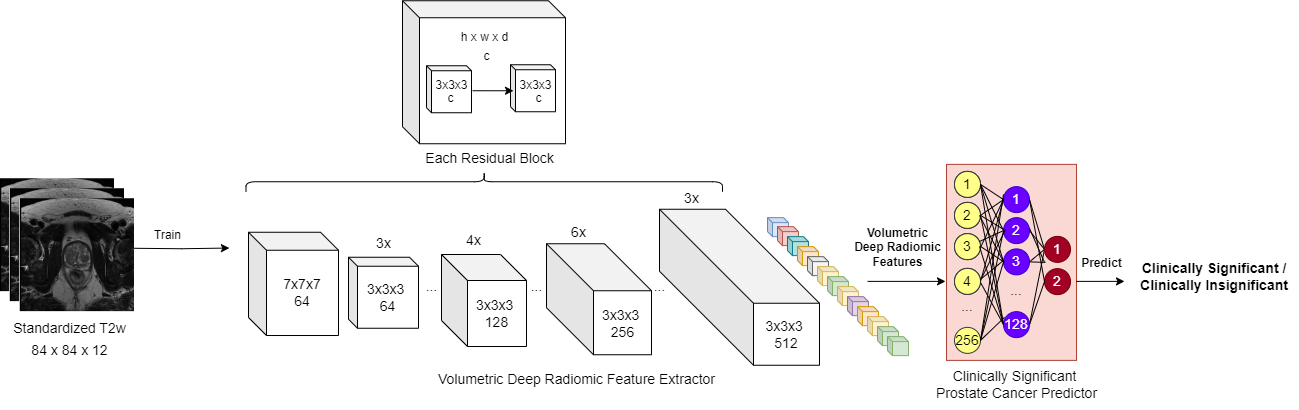}
  \caption{Workflow for predicting clinically significant prostate cancer using volumetric deep radiomic features with model structure adapted from~\cite{tai2023enhancing}.}
  \label{fig:implemented-workflow}
\end{figure*}

For this study, we use the cohort of 200 patient cases obtained from Radboud University Medical Centre (Radboudumc) in the Prostate MRI Reference Center in Nijmegen, The Netherlands~\cite{6729091}. Patients were labelled true for clinically significant prostate cancer if their prostate contained any tissue with Gleason Grade Groups 2-5 according to the International Society of Urological Pathology (Gleason score greater than or equal to 7)~\cite{PLOUSSARD2011291}. Otherwise, the patient was labelled as clinically insignificant prostate cancer. In the cohort, all patients had prostate cancer.

A Siemens MAGNETOM Trio 3.0T machine or a Siemens MAGNETOM Skyra 3.0T machine was used to acquire the patient images. An expert radiologist with over 20 years of experience interpreting prostate MRI was involved in the supervision of the reviewal or acquisition of the MRI~\cite{6729091}. A turbo spin-echo sequence with TR ranging from 3880 to 7434.8 ms (median = 5660 ms) and TE ranging from 101 to 112 ms (median = 104 ms) was used to obtain the T2-weighted acquisitions. The range of the slice thickness was 3 to 4.5 mm (median = 3 mm) and the range of the display field of view was 18 x 18 cm\textsuperscript{2} to 19.2 x 19.2 cm\textsuperscript{2} (median = 19.2 x 19.2 cm\textsuperscript{2}). The range of the in-plane resolution of the acquisitions was 0.3 to 0.6 mm (median = 0.5 mm)~\cite{6729091}.

The workflow shown in Figure~\ref{fig:implemented-workflow} was employed for learning deep radiomic features from the patient cohort, with a predictor to output the associated clinical significance prediction. To assess the efficacy of transfer learning from breast cancer, the performance with the volumetric deep radiomic feature extractor initialized with MONAI weights~\cite{monai} was compared to that initialized with the pretrained breast cancer grade weights (BCa)~\cite{tai2023enhancing}. The pretrained breast cancer grade weights were obtained by training a volumetric deep radiomic feature extractor on MRI images of breast cancer patients to predict the Scarff-Bloom-Richardson (SBR) grade, a metric describing the severity and spread of the breast cancer tumour~\cite{tai2023enhancing}. On the other hand, the MONAI weights were obtained through training on a large-scale 3D medical dataset, 3DSeg-8, a consolidated dataset of images from eight 3D segmentation datasets containing both MRI and CT images~\cite{chen2019med3d}.  

For training, a learning rate of 1e-05 was used along with a weighted random sampler, AdamW optimizer, cross-entropy loss function and cosine annealing learning rate scheduler. The random horizontal flip and random vertical flip image transforms were also included during training. Leave-one-out cross-validation was conducted to obtain the results for each type of weight initialization with the average accuracy, sensitivity, specificity, and F1 score recorded.

\section{Results}
\label{sec:results}
\begin{figure}
    \centering
    \subfloat[]{\includegraphics[width=0.45\linewidth]{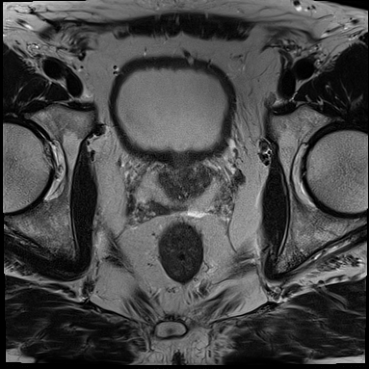}} \hfil
    \subfloat[]{\includegraphics[width=0.45\linewidth]{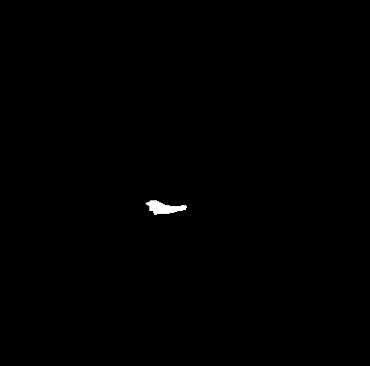}} \hfil \\
    \subfloat[]{\includegraphics[width=0.45\linewidth]{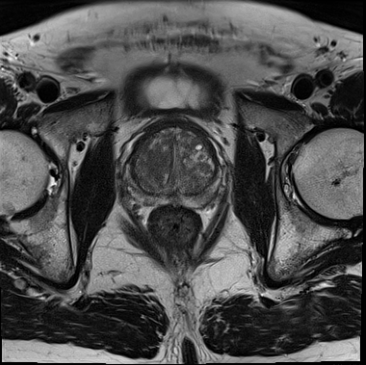}} \hfil
    \subfloat[]{\includegraphics[width=0.45\linewidth]{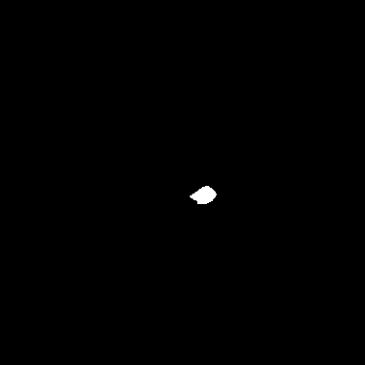}} 
    \caption{Sample T2w image and associated mask for clinically significant (a, b) and clinically insignificant (c, d) prostate cancer.}
    \label{fig:sample-t2w-image}
\end{figure}

As seen in Table~\ref{tab:pca-results}, the workflow initialized using transfer learning from breast cancer obtained a leave-one-out cross-validation accuracy of 97.50\%, over 30\% higher compared to the one initialized with MONAI weights (63.00\%). Using transfer learning also achieved higher sensitivity, specificity, and F1 scores compared to MONAI initialization, with all metrics over 90\%. Figure~\ref{fig:sample-t2w-image} shows sample T2w images for a patient who has clinically significant prostate cancer and a patient that has clinically insignificant prostate cancer. In both of these situations, the model with transfer learning, initialized with BCa weights was able to predict the correct significance class. These promising results highlight the efficacy of cross-domain knowledge transfer; by leveraging pretraining based on another type of cancer (breast cancer), we were able to achieve a significant improvement in performance. Future work includes replicating this ablation study for other types of cancer to further generalize these findings.

\begin{table}
\caption{Accuracy (Acc.), Sensitivity (Sens.), Specificity (Spec.), and F1 score for workflows initialized with MONAI weights and pretrained breast cancer grade (BCa) weights.}
    \centering
    \begin{NiceTabular}{l c c c c}
        \toprule
        \RowStyle{\bfseries}
        Weights & Acc. & Sens. & Spec. & F1 score \\ \midrule
        MONAI~\cite{monai} & 63.00\% & 84.29\% & 51.54\% & 61.46\% \\
        BCa~\cite{tai2023enhancing} & \textbf{97.50\%} & \textbf{94.29\%} & \textbf{99.23\%} & \textbf{96.35\%} \\
    \bottomrule
    \end{NiceTabular}
    \label{tab:pca-results}
\end{table}

{
    \small
    \bibliographystyle{ieeenat_fullname}
    \bibliography{main}
}

\end{document}